\documentclass[fleqn,usenatbib]{mnras}

\usepackage{newtxtext,newtxmath}

\usepackage[T1]{fontenc}
\usepackage{ae,aecompl}

\usepackage{graphicx}	
\usepackage[switch]{lineno}

\usepackage{mathtools} 
\newcommand{\be}{\begin{equation}}
\newcommand{\ee}{\end{equation}}
\newcommand{\bea}{\begin{eqnarray}}
\newcommand{\eea}{\end{eqnarray}}

\title[magnetized GRB afterglows dynamical model]{An energy-conserving dynamical model of GRB afterglows from magnetized forward and reverse shocks}

\author[Q. Chen \& X. W. Liu]{
Qiang Chen,$^{1,2}$\thanks{E-mail: chen@camk.edu.pl}
Xue-Wen Liu$^{2}$
\\
$^{1}$Nicolaus Copernicus Astronomical Center, Polish Academy of Sciences, Bartycka 18, 00-716 Warsaw, Poland\\
$^{2}$Center for Theoretical Physics, College of Physical Science and Technology, Sichuan University, Chengdu, 610064, PR China
}

\date{Accepted 2021 March 31. Received 2021 March 31; in original form 2019 June 27}
\pubyear{2021}

\begin{document}
\label{firstpage}
\pagerange{\pageref{firstpage}--\pageref{lastpage}}
\maketitle

\begin{abstract}

In the dynamical models of gamma-ray burst (GRB) afterglows, the uniform assumption of the shocked region is known as provoking total energy conservation problem.
In this work we consider shocks originating from magnetized ejecta, extend the energy-conserving hydrodynamical model of \cite{1009-9271-7-6-05} to the MHD limit by applying the magnetized jump conditions from \cite{Zhang:2004ie}.
Compared with the non-conservative models, our Lorentz factor of the whole shocked region is larger by a factor $\lesssim\sqrt{2}$. 
The total pressure of the forward shocked region is higher than the reversed shocked region, in the relativistic regime with a factor of about 3 in our interstellar medium (ISM) cases while ejecta magnetization degree $\sigma<1$, and a factor of about 2.4 in the wind cases.  
For $\sigma\le 1$, the non-conservative model loses $32-42$\% of its total energy for ISM cases, and for wind cases $25-38$\%, which happens specifically in the forward shocked region, making the shock synchrotron emission from the forward shock less luminous than expected.
Once the energy conservation problem is fixed, the late time light curves from the forward shock become nearly independent of the ejecta magnetization.
The reverse shocked region doesn't suffer from the energy conservation problem since the changes of the Lorentz factor are recompensed by the changes of the shocked particle number density.
The early light curves from the reverse shock are sensitive to the magnetization of the ejecta, thus are an important probe of the magnetization degree.

\end{abstract}


\section{Introduction}
\label{sec:intro}

Gamma-ray bursts (GRBs) are the most luminous transient sources of electromagnetic radiation in the Universe (for review, see \citealt{Meszaros:2006rc}), and they became essential tools in cosmology \citep{10.1111/j.1365-2966.2009.15456.x,10.1111/j.1365-2966.2010.17197.x,10.1093/mnras/stt1516,Postnikov_2014,WANG20151,DAINOTTI201723,2018AdAst2018E...1D,Dainotti_2018,10.1088/2053-2563/aae15c}, and multi-messenger astronomy \citep{Abbott_2017}. The prompt gamma-ray emission of a GRB lasts for only a few seconds, but the successive afterglow can be observed for months at multiple wavelengths, from the X-rays to the radio, and even in the very-high-energy gamma rays \citep{Arakawa:2019cfc}. Observations of the afterglows are important for locating the host galaxy, for example by measuring the distance from its redshift \citep[e.g.][]{Reichart_1998}, and also in constraining the geometry and the dynamics of the outflow (e.g. \citealt{Sari_1999,Moderski_2000}).

In the standard GRB fireball model (FBM) \citep{1986ApJ...308L..47G,1986ApJ...308L..43P}, the term "fireball" refers to an opaque radiation-plasma whose initial energy is significantly greater than its rest mass.
Initially, the radiation-dominated fireball expands relativistically outwards under its own pressure, and baryons are accelerated with the fireball and converted part of the radiation energy into bulk kinetic energy.
At some point, the fireball transits into the matter-dominated phase, the ejecta then experience coasting, deceleration, and non-relativistic (Newtonian) sub-phases \citep{Piran:1999kx}.
The so-called afterglow originates from the matter-dominated phase where GRB ejecta colliding with the circum-burst medium (CBM), resulting in a forward shock (FS) propagating into the CBM, and potentially also a reverse shock (RS) propagating into the ejecta. As the FS sweeps significant amounts of CBM mass, the shell of the shocked plasma decelerates gradually \citep{10.1093/mnras/258.1.41P}. These relativistic shocks are collisionless, and they can accelerate particles in the stochastic processes \citep[e.g.][]{BLANDFORD19871}, and the resulting non-thermal particles can then release their energy through radiative processes like synchrotron radiation and inverse Compton scattering.

The GRB afterglow is the external explosion, where the evolution of pressure, velocity, etc. behind the shock do not depend significantly on the precise details of the internal explosion, but only on the ejecta energy and the property of the unshocked materials.
So the afterglow can be well described by \emph{self-similar solutions} \citep{doi:10.1063/1.861619}.
Self-similar solutions could reduce the partial differential equations (PDEs) into ordinary differential equations (ODEs), allowing exactly analytic solutions to be found \citep{doi:10.1063/1.1285921,doi:10.1063/1.2174567}.
The first found self-similar solutions are known as Sedov-Taylor solutions (hereafter ST), which focused on the non-relativistic shocks \citep{1959sdmm.book.....S,doi:10.1098/rspa.1950.0049}.
Then \citet{doi:10.1063/1.861619} (hereafter BM) found self-similar solutions for the adiabatic relativistic deceleration phase.
These solutions are derived by solving relativistic shock jump conditions with the assistance of extra equations of setting the divergence of the relativistic energy-momentum tensor equal to zero, and also the equation of particle continuity.
The radiative version of BM self-similar solutions is derived by \citet{Cohen_1998}.
The ST and BM solutions are called the first type self-similar solutions, where the scaling of the radius as a function of time for the problem is determined by dimensional considerations alone \citep{doi:10.1063/1.1285921}.
In comparison for the second type self-similar solutions, the scaling must be found by demanding that the solution passes through a singular point of the equation \citep{doi:10.1063/1.2174567}. 
To cover certain extreme CBM profiles of the non-relativistic shocks, the ST solutions are extended to the second-type self-similar solutions \citep{doi:10.1063/1.858668}.
Similarly, the relativistic BM solutions are also extended to the second-type solutions \citep{doi:10.1063/1.1285921}.

Later, many \emph{(FS) generic dynamical models} have been proposed both for the relativistic and Newtonian phases, either in adiabatic or radiative regimes \citep{Chiang_1999,Piran:1999kx,Huang:1999di,2041-8205-752-1-L8,doi:10.1093/mnras/stt872,2019arXiv191107350Z}.
The idea is based primarily on solving the shock jump conditions (local conservation of energy and momentum) with the help of an additional total energy conservation equation in each shell region of the afterglow (global conservation of energy).
In this paper, we refer to these generic models as \emph{(global) energy-conserving models}.

As for the double shock system including both forward and reverse shocks (FS-RS), the dynamical models can be developed analogously to the energy-conserving models of the single FS, but historically another easier approach was proposed, even earlier than the FS generic dynamical models.
The FS-RS involves two individual sets of shock jump conditions, and therefore two additional equations are needed to solve them.
The first handy condition is to assume that FS and RS have equal velocity, which seems to be reasonable.
If further assume that the pressures of the FS and RS are uniform and equal, the shock jump conditions are solvable, without invoking constraints on the global energy or momentum. 
This approach was explored analytically \citep{1995ApJ...455L.143S} and numerically \citep{2000ApJ...545..807K}.
In this paper, we refer to this method as the \emph{jump condition model}.

However, the assumption of uniform pressure in the whole shocked region of the FS-RS system has been questioned for violating energy conservation \citep{2006ApJ...651L...1B}.  
Especially in the case of long-lived RS, the energy conservation problem of the jump condition model cannot be ignored \citep{Uhm_2011,Uhm_2012, Geng_2016}.   
A \emph{mechanical model} has been developed by applying conservation laws for the energy-momentum tensor and the mass flux to the blast region between the FS and the RS \citep{2006ApJ...651L...1B}.  
The uniform pressure assumption has been dropped out, meanwhile, while the assumption of uniform Lorentz factor (velocity) has been preserved.
Indeed, the uniform pressure assumption and the uniform Lorentz factor assumption cannot be adopted at the same time.
The mechanical model shows the pressure of the FS shocked region is higher than the RS shocked region by as much as a factor of 3 in some of their chosen example.

While the jump condition model has an energy-conserving problem, and on the other hand the mechanical model is complicated, extending the energy-conserving models to the FS-RS becomes another option.
The difference between the jump condition model and the FS-RS energy-conserving model is that the latter replaced the uniform pressure assumption with the global energy conservation equation, while the uniform Lorentz factor assumption is preserved in both models.
This makes the energy-conserving model more complex than the jump condition model.
The FS-RS energy-conserving models are also been developed for shocks from the hydrodynamical ejecta \citep{1009-9271-7-6-05,doi:10.1093/mnras/stt872,Geng_2014}.

Regarding afterglows from the magnetized ejecta, there are several available frameworks to choose from.
The first option is to extend the standard FBM, where most energy produced by the central source is carried by the bulk motion of baryons \citep{Piran:2004ba}, to the MHD limit.
E.g., embedded with the magnetized shock jump conditions, by equalizing the total pressure of the RS shocked region, including both the thermal pressure and the magnetic pressure, with the thermal pressure of FS shocked region, the jump condition model has been extended to the magnetized cases \citep{2004A&A...424..477F,Fan:2004fe,Zhang:2004ie}.
The RS won't always emerge from the magnetized ejecta, above some critical value, it will be suppressed by the strong field \citep{2009A&A...494..879M}.
While some models claim the ejecta magnetization can reach high values, even to the order $\sigma\sim 100$s \citep{Zhang:2004ie}, the others estimate that the ejecta with magnetization $\sigma\gtrsim 1$ are not crossed by an RS for a large fraction of the parameter space relevant to the GRB flows \citep{Giannios:2007tu}.
So the existence condition of the RS is worthy of continuing investigations \citep{Mizuno_2008,10.1111/j.1365-2966.2010.17696.x}.
Because the magnetized jump condition model adopts the uniform pressure assumption, the total energy conservation is a remained problem which we are trying to resolve in this paper.

An alternative option is the \emph{electro-magnetic model} (EMM), where the bulk energy of the ejecta flow is carried by the magnetic field \citep{2003astro.ph.12347L,2004cosp...35..237L, Lyutikov_2006}.
Under the framework of EMM, the afterglow dynamics can be derived starting from the total energy conservation, by injecting an arbitrary fraction of the electromagnetic energy ($\sigma\gg 1$) into the FS.
Usually, the early afterglows are dominated by the RS, however, the RS is intrinsically absent in the EMM, yet the early-time afterglows in the EMM are still found brighter than in the standard FBM \citep{GenetF.2006,GenetF.2007}.

Meanwhile, with the rapid development in computational science, \emph{numerical simulations} became another alternative option in studying the magnetized GRB shocks. \citet{2009A&A...494..879M} performed 1D simulations with magnetized ejecta and concluded that the RS is weak or absent in ejecta characterized by $\sigma\gtrsim 1$.  The particle-in-cell (PIC) simulations show that only weakly magnetized shocks are efficient particle accelerators \citep{2011ApJ...726...75S,2013ApJ...771...54S,Sironi2015}, and these discoveries will enhance our understanding of the physical mechanism.

Once we have the results from the dynamical models, it's possible to undertake afterglow emission analysis.
For convenience, usually assume a certain fraction $\epsilon_e$ of the thermal energy in the shocked region goes into the electrons, then the electrons are accelerated in the shocks to a power-law distribution \citep{Sari:1997qe}.
Ejecta magnetization has an obvious impact on both the RS dynamics and its synchrotron emission output.  
For low magnetization degrees, the luminosity of the RS increases steadily with increasing $\sigma$, reaches the highest levels for $\sigma \sim 0.1-1$, and then decreases for even higher $\sigma$.
Another signature of the high-$\sigma$ is that the optical RS peak is broadened \citep{2004A&A...424..477F,Zhang:2004ie}. 

Early optical flashes have been postulated as signals produced from the RS \citep{10.1046/j.1365-8711.1999.02800.x,Sari_Piran_1999,Kobayashi_zhang_2003}, however, only a small fraction of GRBs exhibit a bright optical flash \citep{Yost_2007, Melandri_2008, Klotz_2009,Rykoff_2009}. 
Radio flares have also been attributed to the  RS \citep{Kulkarni_1999,Berger_2003,Chevalier_2004}.  
In some cases, even though an optical flash is not detected, radio observations by very sensitive radio telescopes, e.g. the upgraded Jansky Very Large Array (JVLA), can capture the signals from the  RS \citep{Mundell_2007,Laskar_2013, Laskar_2013_1, Kopa__2015, Laskar_2016, Alexander_2017,Laskar_2018_1, Laskar_2018, Laskar_2019, Laskar_2019_1}.

In this paper, we build a new energy-conserving dynamical model, by extending the hydrodynamical FS-RS model proposed by \cite{1009-9271-7-6-05} to the MHD limit with the magnetic jump conditions conventions of \citet{Zhang:2004ie}.
Section \ref{sec:conf} presents the model assumptions of our GRB scenario, including the model parameters (Sec. \ref{sec:basic}), magnetized jump conditions with numerical solutions under the assumption of uniform pressure, and Lorentz factor, where the conflict between the numerical solution and the uniform Lorentz factor assumption reveals the origin of the energy conservation problem (Sec. \ref{sec_origin}), and the existence conditions for the RS in the parameter space of initial ejecta Lorentz factor and magnetization ($\eta$-$\sigma$) with their numerical solutions (Sec. \ref{sec_RSexist}). 
Section \ref{sec:model} is the description of our new energy conserving dynamical model. In Section \ref{sec:result} we present the numerical results of the dynamics of the new model (Sec. \ref{sec:result_dyn}), the conservation of energy (Sec. \ref{sec_conser}), and the pressure  balance  between the shocked regions (Sec. \ref{sec:pressure}).  Section \ref{sec:syn} presents the synchrotron radiation light curves based on the dynamical results.  
We present the magnetic energy fraction for radiation in the shocked ejecta region (Sec. \ref{sec:frac}).   
We test both ISM and wind, adiabatic and radiative cases in the optical (Sec. \ref{sec:optic}) and radio (Sec. \ref{sec:radio}) observational bands.  We also test our model by comparing it to the multiple wavelength observational light curves of GRB181201A (Sec. \ref{sec:GRB181201A}).  And Section \ref{sec:conc} is devoted to the discussion and conclusions.

\section{Model assumptions and parameters}
\label{sec:conf}

In this section, we are going to describe the essential assumptions and parameters for building a magnetized afterglow model.
Oftentimes these assumptions may be analogous to the hydrodynamical afterglow.
Nevertheless, the ejecta-related parameters and relations that are commonly adopted in the hydrodynamical afterglow model should be most probably updated.
The basic assumptions and parameters are described in Section \ref{sec:basic}.
The magnetized relativistic shock jump conditions are described in Section \ref{sec_origin}, which should also be the very central equations that need to be solved in the rest of the paper.
Last but not least, one of the major differences between the hydrodynamical and the magnetized afterglow is that the magnetization of the ejecta can suppress the existence of the RS, which if true will invalid our model assumption.
We will describe this in Section \ref{sec_RSexist}.

\subsection{Basic assumptions and parameters}
\label{sec:basic}

\begin{figure}
\centering
\includegraphics[width=0.49\textwidth]{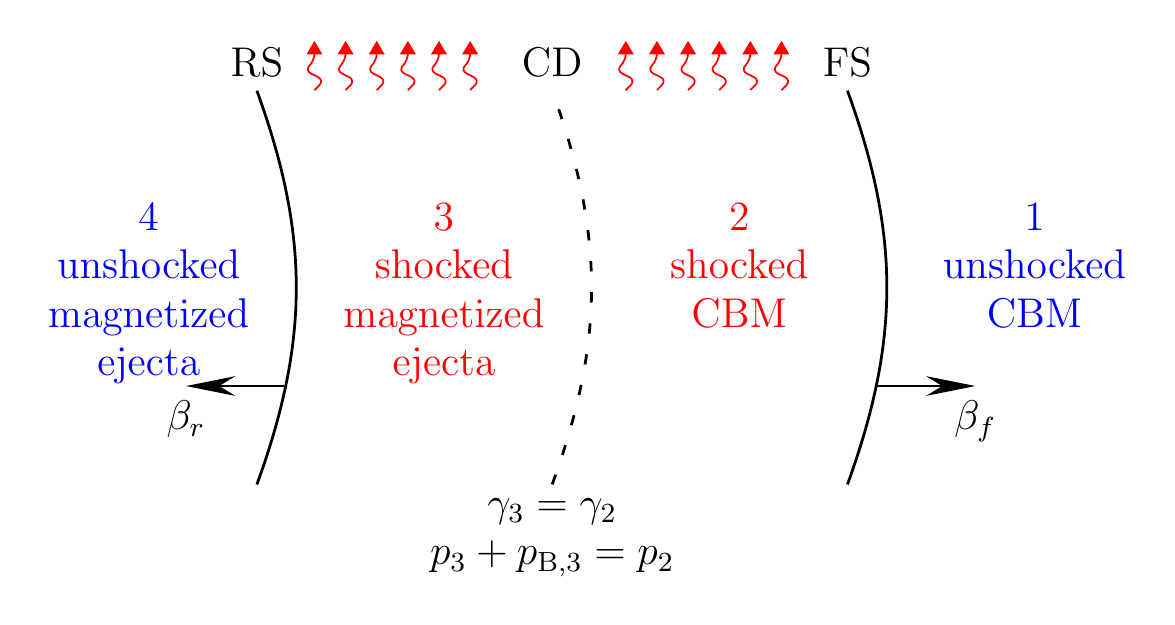}
\caption{
Illustration of the GRB afterglow components before the RS crossing time. 
The cold magnetized ejecta (region 4) is clashing with cold hydrodynamical circum-burst medium (region 1). 
From contact discontinuity (CD) launches a forward shock (FS) and a reverse shock (RS) (if exists).
For explicit, here velocities $\beta_f$ and $\beta_r$ are measured in the rest frame of the blast wave (shocked region).
Regions behind the shocks are shocked and emitting energy away.
The usual assumptions of the uniform Lorentz factor and uniform pressure are $\gamma_3=\gamma_2$ and $p_3+p_{B,3}=p_2$.
The uniform Lorentz factor assumption is more or less reasonable, but these two assumptions will conflict with each other, and cause an energy conservation problem.
}
\label{fig_shell}
\end{figure}

Consider a cold ejecta shell with mass $m_{\rm ej}$, initial width $\Delta_0$, and initial Lorentz factor $\eta$, that collides with the cold CBM at the radius $R_0$ from the source engine, where a double shock structure is potential to be launched. 
The CBM and the ejecta shell are separated by the  FS, CD, and RS into four regions (see illustration in Figure \ref{fig_shell}): (1) the unshocked CBM, (2) the shocked CBM, (3) the shocked ejecta, and (4) the unshocked ejecta \citep{1994ApJ...422..248K}.
We use single subscript indexes to indicate parameters in different regions, e.g. $\gamma_2$ denotes Lorentz factor in Region 2, $p_f$ denotes pressure at the  FS front.
We use double indexes to describe the relative value between two regions, e.g., $\gamma_{34}$ means the Lorentz factor of Region 3 in the rest frame of Region 4, and $u_{3r}$ means the four-velocity of Region 3 in the rest frame of the RS.

We assume that the ejecta are magnetized, and their magnetization parameter is approximated  with the Poynting-to-kinetic flux ratio $\sigma_4 = U_B/E_k$ \citep{Zhang:2004ie,Giannios:2007tu}.
In this paper we assume that magnetization of the CBM is negligible, and that $\sigma_4$ is not evolving with radius $R$, therefore, we omit the subscript. The initial total energy of the system equals the initial energy contained in the ejecta, so that $E_0 = E_k(1+\sigma) = \eta m_{\rm ej} c^2 (1+\sigma)$ where $c$ is the speed of light. The co-moving particle number density is decreasing with radius $R$ as $n_4 = m_{\rm ej}/(4\pi R^2\Delta \eta m_p) = E_0/[4\pi R^2\Delta\eta^2 m_pc^2 (1+\sigma)]$, where $\Delta = \max[\Delta_0,R/\eta^2]$ is the shell width that takes into account the spreading effect \citep{1995ApJ...455L.143S}.  The CBM number density is parameterized as $n_1(R) = AR^{-k}$, where for ISM case, $k=0$, and $A = n_0$, with the typical value $n_0=1 {\rm cm}^{-3}$. For wind case $0<k<4$, $A=3\times 10^{35} A_\ast{\rm cm}^{-1}$, where $A_\ast$ is a dimensionless parameter \citep{10.1093/mnras/258.1.41P,1994ApJ...422..248K,Chevalier_2000,Meszaros:2006rc}.

The initial fireball size is $R_i\approx cT$, where $T$ is the duration of the burst.
The initial fireball size roughly equals the initial shell width $\Delta_0$.
The expansion of the fireball transits from the radiation-dominated phase to the matter-dominated phase at radius $R_\eta = R_i\eta$ \citep{10.1093/mnras/263.4.861,Piran:1999kx}.
We choose the transition radius as the onset radius of the RS, i.e. $R_0 = R_\eta$.
Since the shell width is relatively thin compared to the shock radius, i.e. $\Delta\ll R$, it can be assumed that the FS, CD, and RS  have the same radii.  To the on-axis observer, the shock radius $R$ equation from the observer time is \citep{Meszaros:2006rc}:
\be
\frac{{\rm d}R}{{\rm d}t} = \frac{\beta_2 c}{1-\beta_2}\frac{1}{1+z}\,,  
\ee
where $z$ is the cosmological redshift, and $\beta_2$ is the dimensionless velocity of region 2.  In this paper, all velocities are measured in the rest frame of the source, unless stating otherwise.

The mass of Region 2 is made of the CBM material collected by  the FS , and the mass of Region 3 is made of the GRB ejecta that cross the RS \citep{Huang:1999di,1009-9271-7-6-05,doi:10.1093/mnras/stt872}:
\bea  \label{eq_dm2}  {\rm d}m_2 &=& 4\pi R^2 n_1m_p\,{\rm d}R\,,  \\  \label{eq_dm3}  {\rm d}m_3 &=& 4\pi R^2 n_4m_p \gamma_4\left(\frac{\beta_4-\beta_r}{\beta_r}\right)\,{\rm d}R\,,  \eea
where $\beta_r$ is the dimensionless velocity of the RS.

\subsection{Magnetized jump conditions}
\label{sec_origin}

\begin{figure}
\centering
\includegraphics[width=0.49\textwidth]{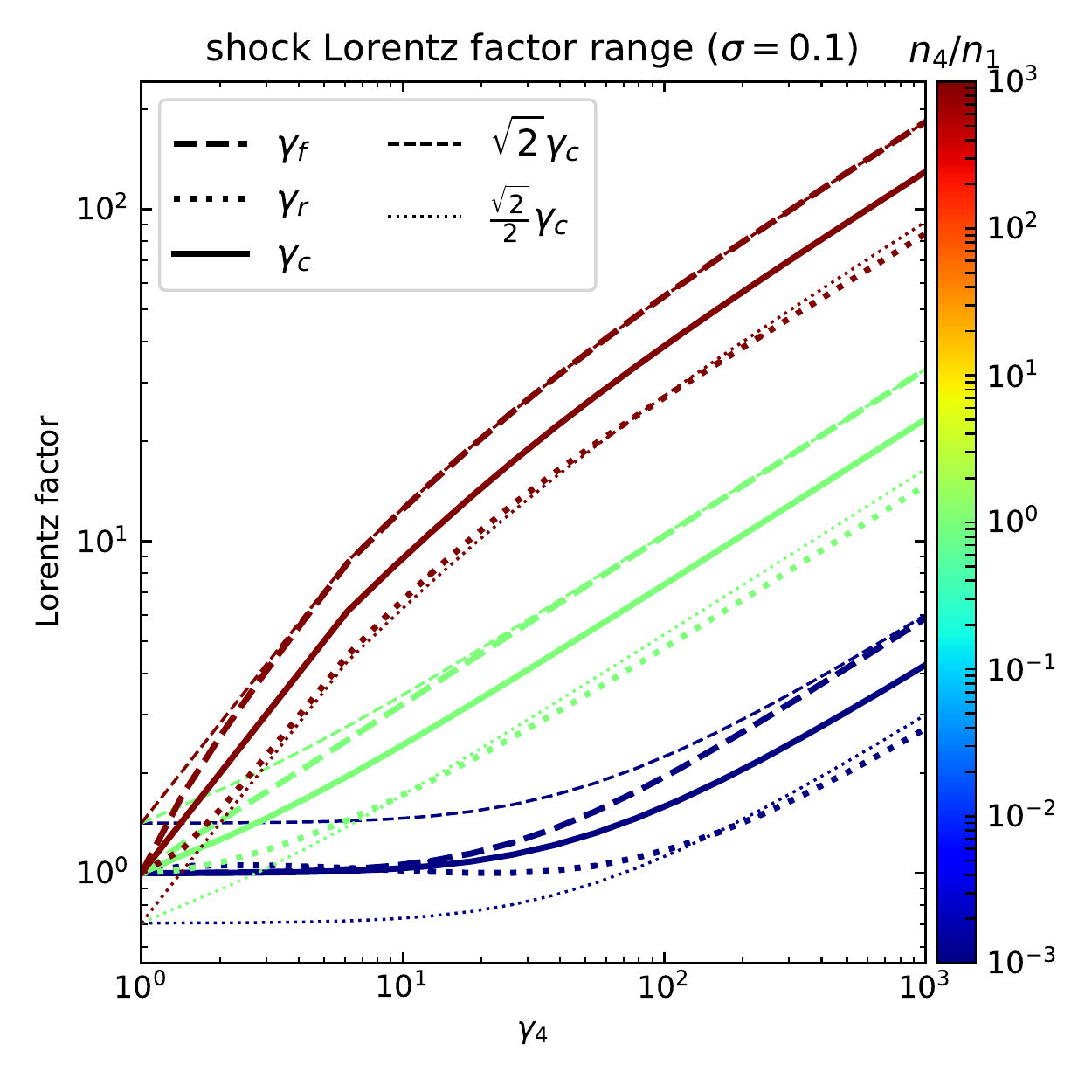}
\caption{
Shock Lorentz factors relation for the case $\sigma=0.1$ derived from the magnetized shock jump conditions with the help of uniform Lorentz factor and pressure assumption.
The colorbar is the particle number density ratio $n_4/n_1$.
A clear relation is shown as: $\gamma_r  <\gamma_c  < \gamma_f$.
In the ultra-relativist limit $\sqrt{2}\gamma_r \approx \gamma_c\simeq\sqrt{2}/2\gamma_f$.
This gives the upper and lower boundaries of the values of $\gamma_3$ and $\gamma_2$, which are apparently not to be equal.
This implies the simplification of $\gamma_2=\gamma_3=\gamma_c$ involves with an error by a factor $\lesssim\sqrt{2}$. 
The conclusion is that the uniform pressure assumption could invalid the uniform Lorentz factor assumption.
}
\label{fig_Lorentz}
\end{figure}

The GRB afterglow can be described by the relativistic shock jump conditions \citep{doi:10.1063/1.861619}, which are derived from conservation laws across each shock front.  The shock jump conditions of the afterglow from magnetized ejecta are \citep{Zhang:2004ie}:
\bea  \frac{e_2}{n_2m_pc^2}&=&\gamma_{21}-1\,,  \label{eq_e2n2}  \\
\frac{n_2}{n_1}&=&\frac{\hat\gamma_2\gamma_{21}+1}{\hat\gamma_2-1}\,,  \\
\frac{e_3}{n_3m_pc^2}&=& (\gamma_{34}-1)f_a\,,  \label{eq_e3n3}  \\
\frac{n_3}{n_4}&=&\frac{\hat\gamma_{3}\gamma_{34}+1}{\hat\gamma_{3}-1}f_b\,,  \label{eq_n3n4}  \eea
with  parameters $f_a$, $f_b$ and $f_c$ defined as:
\bea  \label{fa}  f_a&=&1-\frac{\gamma_{34}+1}{2[u_{3r}^2\gamma_{34}+u_{3r}(u_{3r}^2+1)^{1/2}(\gamma_{34}^2-1)^{1/2}]}\sigma\,,  \\
f_b&=&\left(\gamma_{34}+\sqrt{(u_{3r}^2+1)(\gamma_{34}^2-1)}/u_{3r}\right)\frac{\hat\gamma_{3}-1}{\hat\gamma_{3}\gamma_{34}+1}\,,  \\
f_c&=&1+\frac{p_{B,3}}{p_3} = 1+\frac{1}{2(\hat\gamma_{3}-1)}\frac{n_3}{n_4}\left(\frac{e_3}{n_3m_pc^2}\right)^{-1}\sigma\,,  \eea
where $e_3$ is the energy density, $p_{B,3} = B_3^2/(8\pi)$ and $p_3$ are the co-moving magnetic pressure and thermal pressure respectively, and $\hat\gamma_i$ is the adiabatic index \footnote{For relativistic ideal gas, let $\Theta = p/nm_pc^2$ be the dimensionless temperature, where $p$ is the thermal pressure, $n$ is the particle number density. The adiabatic index $\hat\gamma$ is then $\hat\gamma(\Theta) = 1 + \Theta/[3\Theta + h(\Theta) - 1]$, where $h(\Theta) = K_1(1/\Theta)/K_2(1/\Theta)$ and $K_n(x)$ are the modified Bessel functions of the second kind.} of the plasma in Region "i".

The unknown parameters are $n_2$, $n_3$, $\gamma_2$, $\gamma_3$, $e_2$ and $e_3$ (or equivalently $p_2$, $p_3$), which means that the magnetized jump conditions with inputs $f\equiv n_4/n_1$\citep{1995ApJ...455L.143S}, $\gamma_4=\eta$,  $\gamma_1$ (which is unity as default), and $\sigma$ are not sufficient to provide a unique solution.
Additional equations are needed, like usually the assumption of the uniform velocity and pressure  across the shocked regions \citep{Zhang:2004ie}: 
\bea  
\gamma_2 &=& \gamma_3\,,  
\label{eq_equity_Lorentz}  
\\
p_2 &=& p_3+p_{B,3}\,.  
\label{eq_equity_pressure}  
\eea
The above equations extend the values at the CD to the whole shocked region, in reality, the shocked regions are not uniform.
Take the Lorentz factor as an example, the uniform Lorentz factor assumption adopts $\gamma_2 =\gamma_c =\gamma_3$ instead of real relation $\gamma_r < \gamma_3 <\gamma_c < \gamma_2 < \gamma_f$.
To estimate the span between $\gamma_r$ and $\gamma_f$, in a hydrodynamical afterglow system \citep{doi:10.1063/1.861619}:
\bea  \gamma_{1f}^2  &=&\frac{(\gamma_{21}+1)[\hat\gamma_2(\gamma_{21}-1)+1]^2}{\hat\gamma_2(2-\hat\gamma_2)(\gamma_{21}-1)+2} \,,  \\
\gamma_{4r}^2  &=&\frac{(\gamma_{34}+1)[\hat\gamma_3(\gamma_{34}-1)+1]^2}{\hat\gamma_3(2-\hat\gamma_3)(\gamma_{34}-1)+2} \,.  \eea
In the ultra-relativistic limit, $\gamma_{1f}\simeq \sqrt{2}\gamma_{21}$ and $\gamma_{4r} \simeq \sqrt{2}\gamma_{34}$.  

In the MHD limit, to derive an analytical expression for $\gamma_{4r}$ is not practical, since we need to know $u_{3r}(\gamma_{34},\sigma)$, which is a root of a six-order equation \citep{Zhang:2004ie}.
We made an exercise to numerically solve the magnetized jump conditions with the assumption of uniform velocity and pressure, i.e. the jump condition model \citep{Zhang:2004ie}.
We show an example case of $\sigma=0.1$ in Figure \ref{fig_Lorentz}.
In the ultra-relativistic limit $\sqrt{2}\gamma_r \approx \gamma_c \simeq \gamma_f/\sqrt{2}$ gives the upper and lower limits of $\gamma_2$ and $\gamma_3$, which are not to be equal, with error factor as large as $\lesssim\sqrt{2}$.
This means the adoption of the uniform pressure assumption invalids the uniform Lorentz factor assumption, and these two assumptions cannot be adopted at the same time.
The conflict between the two assumptions in the jump condition model will inevitably lead to a violation of energy conservation, so a total energy-conserving model is needed to resolve the problem.

\subsection{Condition for the existence of the reverse shock}
\label{sec_RSexist}

\begin{figure*}
\centering
\includegraphics[width=0.99\textwidth]{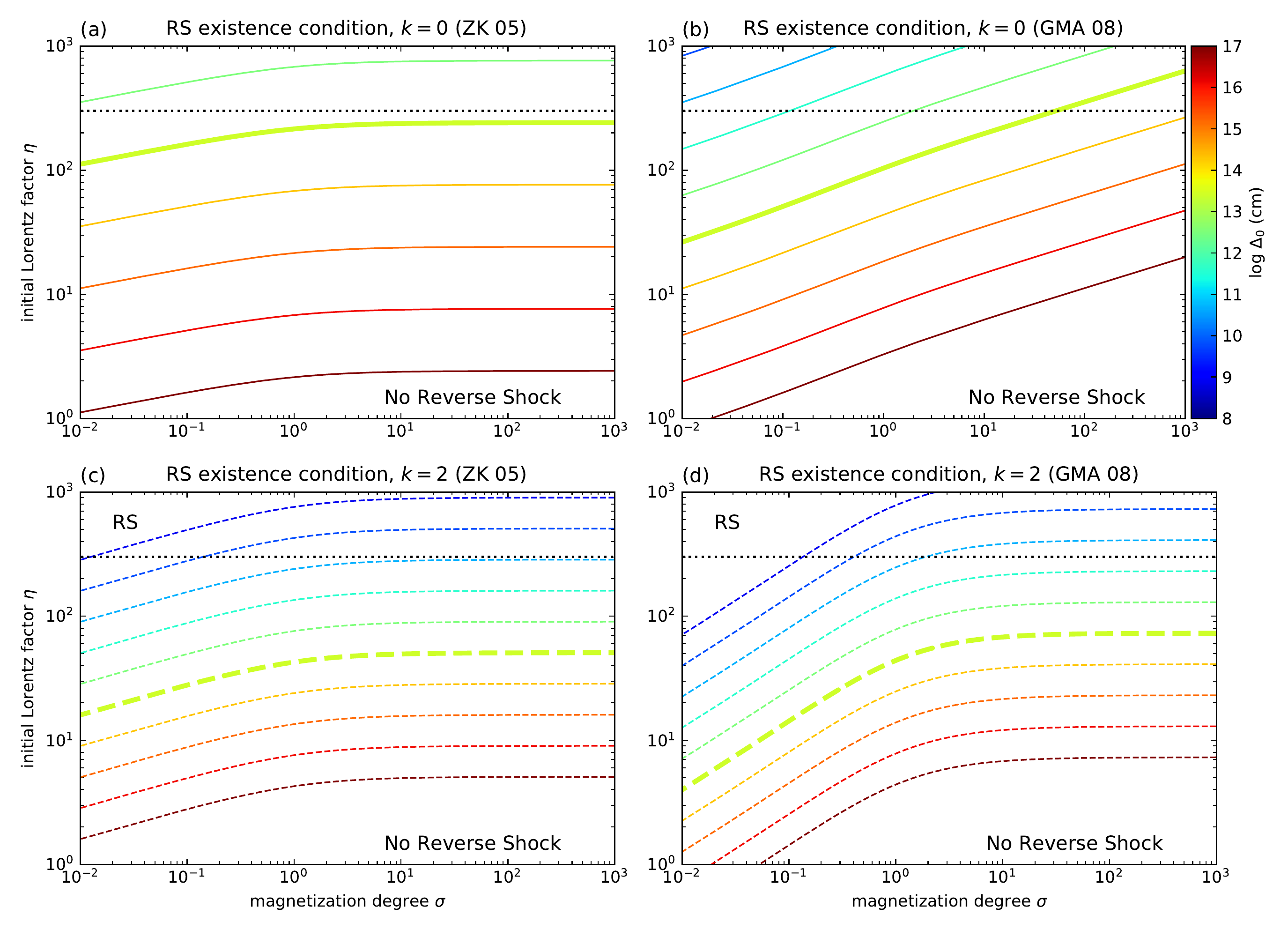}
\caption{
Existence conditions of the reverse shock in $\eta-\sigma$ parameter space.
The solid and dashed lines represent the ISM and wind cases respectively. 
Colors are representing different initial shell widths $\Delta_0$. 
In left column are ISM ($k=0$) and wind ($k=2$) cases based on the ZK 05 approach \citep{Zhang:2004ie}, where we plot the zero contour lines of the pressure balance $\sigma < (8/3)(n_1/n_4)\gamma_4^2$ as the RS existence condition. 
The right column is based on the GMA 08 approach \citep{Giannios:2007tu}, where the radius relation $R_{\rm c} > R_{\rm rs}$ is interpreted as the RS existence condition.
The wind case is more tolerant of a high $\sigma$ in the high Lorentz factors region. 
Smaller shell width is more difficult for ISM to generate RS.
Based on GMA 08, for a specific case with $\Delta_0 = 10^{13}$ cm (ultra-thick lines) and $\eta= 300$ (dotted horizontal line), the RS in the ISM case requires $\sigma \lesssim 50$, while the wind case has no apparent limit.
}
\label{fig_exi}
\end{figure*}

The RS may not always emerge in GRBs with significantly magnetized ejecta \citep{Zhang:2004ie,Giannios:2007tu,10.1111/j.1365-2966.2010.17696.x}.  We use two different approaches to give an insight into adopting proper $\sigma$ values in our coming sections.

One RS existence condition was derived by demanding the ram pressure of the shocked CBM exceeds the magnetic pressure of the ejecta, i.e. $(4/3)\eta^2n_1m_pc^2 > B_4^2/8\pi$, hence $\sigma < (8/3)(n_1/n_4)\eta^2$ \citep{Zhang:2004ie} (ZK 05).  For the ISM case, they derived an explicit constraint as $\sigma <100(\eta/300)^4(T/10{\rm s})^{3/2}(E/10^{52}{\rm ergs})^{-1/2}$. In their chosen cases, RS can survive in high-$\sigma$ regimes when $\sigma$ is less than a few tens or a few hundreds.

A different approach is applied in the final stage of the dynamics evolution at the moment of shell crossing by a condition $R_{\rm c} > R_{\rm rs}$, where $R_{\rm c} \simeq \gamma_4^2\left(\sqrt{(1+\sigma)/\sigma}-1\right)\Delta_0$ is the \emph{MHD contact radius}, and $R_{\rm rs} \simeq (R_{\rm s}\tilde{R}_{\rm dec}^3)^{1/4}/\sqrt{1+\sigma}$ is the \emph{RS crossing radius} with $\tilde{R}_{\rm dec}$ the hydrodynamic \emph{deceleration radius} and $R_{\rm s} = \gamma_4^2\Delta_0$ the \emph{shell spreading radius} \citep{Giannios:2007tu} (GMA 08).
In the ISM case, the RS existence condition is $\xi^3 < (1+\sigma)\left[1+2\sigma-2\sqrt{\sigma(1+\sigma)}\right]/\sigma$ with $\xi = \sqrt{\tilde{R}_{\rm dec} / R_{\rm s}}$, $\tilde{R}_{\rm dec} = (3E/4\pi n_{\rm e}\gamma_4^2m_pc^2)^{1/3}$.
In the wind case, the condition is $\xi_w < (1+\sigma)\left[\sqrt{(1+\sigma)/\sigma}-1\right]$ with $\xi_w = \sqrt{\tilde{R}_{\rm dec,w} / R_{\rm s}}$ and $\tilde{R}_{\rm dec,w} = E/4\pi A\gamma_4^2m_pc^2$.  
This approach was followed and confirmed by \citet{2009A&A...494..879M} using ultrahigh-resolution, one-dimensional, relativistic MHD simulations, who found that the onset of the RS emission is strongly dependent on the magnetization. The RS is typically weak or absent for ejecta characterized by $\sigma \lesssim 1$. 

We compare these two approaches in Figure \ref{fig_exi} with both ISM and wind cases in the initial parameter space of $\eta-\sigma$, where the value of the initial shell width $\Delta_0$ is also taken into account. 
The two approaches result in similar trends, wherewith larger shell width $\Delta_0$ and smaller $\sigma$ enabling the RS for smaller $\eta$.
The GMA 08 approach is stricter than ZK 05, e.g. considering a specific setup with $\eta = 300$ and $\Delta_0=10^{13}$ cm, for the ISM case, ZK 05 predicts that RS should exist for $\sigma\gg 1$, while GMA 08 limits to $\sigma \lesssim 50$.
Because ZK 05 approach assumes the pressure balance at the CD, which is the very assumption that is questioned for the energy conservation problem.
If however being updated with proper pressure relations, the accuracy of the ZK 05 approach may likely improve.

Simulations suggest that high magnetization with $\sigma> 1$ may prevent the forming of RS or result in the inefficient particle acceleration \citep{2009A&A...494..879M,2011ApJ...726...75S}.
The ejecta $\sigma$ are usually be set below unity, since the magnetic field may be significantly reduced by the magnetic dissipation before the afterglow stage \citep{Zhang_2010,Gao_2015,Deng__2016,Deng_2017}.
Due to these reasons current works are more focused on the mild magnetic cases \citep{Mizuno_2008,10.1111/j.1365-2966.2012.20473.x,Lan_2016}.  In the coming sections, we only focus on the $\sigma\lesssim 1$ cases.

\section{Energy Conserving Generic Dynamical Model}
\label{sec:model}

\subsection{Before the RS crossing time}

To resolve the energy conservation problem, we replace the assumption of pressure uniformity  (Eq. \ref{eq_equity_pressure}) with a conservation equation of the total energy. 
Like the mechanical model, we preserve the assumption of the Lorentz factor uniformity  (Eq. \ref{eq_equity_Lorentz}) for simplicity. 
Our generic dynamical model is based on the conservation of the combined total energies of regions 2, 3, and 4 (not including the rest mass energy, same hereafter). 
The gas crossing each shock is heated, and a fraction of $\epsilon$ of the internal energy $U$ is assumed to be radiated away.
Though in reality $\epsilon$ evolves between 1 and 0 \citep{10.1046/j.1365-8711.1998.01681.x,Dai_1999, Feng_2002}, we adopt a constant value for simplicity.
The total energy conservation is \citep{Panaitescu:1998km,Huang:1999di,1009-9271-7-6-05}:
\be\label{eq_tot_ene_cons} 
{\rm d}(E_2 + E_3 + E_4) = - \epsilon_2\gamma_2U_2 \frac{{\rm d}m_2}{m_2} - \epsilon_3\gamma_3U_3 \frac{{\rm d}m_3}{m_3} \,, 
\ee
where $m_2$ and $m_3$ are the masses of region 2 and 3.
The total kinetic energies are expressed in the source frame as:
\bea E_2 &=& (\gamma_2-1)m_2c^2 + (1-\epsilon_2)\gamma_2U_2\,, \\
E_3 &=& (\gamma_3-1)m_3c^2 + (1-\epsilon_3)\gamma_3U_3 + \gamma_3U_{B,3}\,, \\
\label{eq_E3}
E_4 &=& (\gamma_4-1)(1+\sigma)m_4c^2\,, \eea
where $m_4 = m_{\rm ej} - m_3$ is the total mass of the unshocked ejecta.  The co-moving internal energies are inferred from jump conditions, with internal energy in region 3  multiplied by  factor $f_a$ \citep{Zhang:2004ie}:
\bea 
U_2 &=& (\gamma_{2}-1)m_2c^2\,,\\
U_3 &=& f_a(\gamma_{34}-1)m_3c^2\,,
\label{eq_U3}
\eea
where we apply $\gamma_{21}\simeq \gamma_2$ to the  FS,  and $U_{B,3}$ is the magnetic energy of the shocked ejecta.  The relation between the magnetic and internal energies of the shocked ejecta is:
\be
\frac{U_{\rm B,3}}{U_3} = \frac{e_{B,3}}{e_3} = (\hat\gamma_3-1)\frac{p_{B,3}}{p_3} \equiv (\hat\gamma_3-1)(f_c-1)\,. 
\label{eq_eb3e3} 
\ee
With this, the total energy of region 2 becomes:
\be
E_2 = (\gamma_2-1)\left[1 + (1-\epsilon_2)\gamma_2\right]m_2c^2\,.
\label{eq_E2} 
\ee
The total energy of regions 3 can be written as:
\be
E_3 = \left\{\gamma_3-1 + \left[1-\epsilon_3 + (\hat\gamma_3-1)(f_c-1)\right]f_a\gamma_3(\gamma_{34}-1)\right\}m_3c^2\,.
\ee
We now substitute the expressions for $E_2$, $E_3$ and $E_4$ with the assumption of Lorentz factor uniformity to Equation (\ref{eq_tot_ene_cons}).  We have verified numerically that the derivative needs to be applied to parameters $\gamma_2$, $f_a$ and $f_c$, while the terms involving ${\rm d}\hat\gamma_3$ can be omitted.  We use the identity ${\rm d}\gamma_{34} = \gamma_4[1-(\beta_4/\beta_2)]\,{\rm d}\gamma_2$. 
The equation that we obtain is
\be
Q\,{\rm d}\gamma_2 + P_2\,{\rm d}m_2 + P_3\,{\rm d}m_3 + W_a\,{\rm d}f_a + W_c\,{\rm d}f_c = 0\,,
\ee
with the following functions:
\bea Q &\equiv& \left[2\gamma_2-\epsilon_2(2\gamma_2-1)\right]m_2 + m_3+ \left(2\gamma_{34}-\frac{u_4}{u_2}-1\right)\nonumber
\\
  &&\times\left[1-\epsilon_3+(\hat\gamma_3-1)(f_c-1)\right]f_am_3\,,
\\
P_2 &\equiv& \gamma_2^2-1\,, \\
P_3 &\equiv& \gamma_2-1 + \left[(\hat\gamma_3-1)(f_c-1)+1\right]f_a\gamma_2(\gamma_{34}-1)\nonumber
\\
  &&-(\gamma_4-1)(1+\sigma)\,, 
\\
W_a &\equiv& \left[1-\epsilon_3+(\hat\gamma_3-1)(f_c-1)\right]\gamma_2(\gamma_{34}-1)m_3\,, \\
W_c &\equiv& (\hat\gamma_3-1)f_a\gamma_2(\gamma_{34}-1)m_3\,. \eea
The final form of our dynamics equation is \citep{2020past.conf..201C}:
\be
\frac{{\rm d}\gamma_2}{{\rm d}R} 
=-\left(Q + W_a\,\frac{{\rm d}f_a}{{\rm d}\gamma_2} + W_c\,\frac{{\rm d}f_c}{{\rm d}\gamma_2}\right)^{-1}
  \left(P_2\,\frac{{\rm d}m_2}{{\rm d}R} + P_3\,\frac{{\rm d}m_3}{{\rm d}R}\right). \label{eq_result} 
\ee
This equation is integrated numerically, starting from $R_0$, using the 4-th order Runga-Kutte method, with the ${\rm d}m_2/{\rm d}R$ and ${\rm d}m_3/{\rm d}R$ derivatives substituted from Eqs. (\ref{eq_dm2}-\ref{eq_dm3}).  The derivatives ${\rm d}f_a/{\rm d}\gamma_2$ and ${\rm d}f_c/{\rm d}\gamma_2$  are evaluated implicitly.  For $\sigma\ll 1$, $f_a$ and $f_c$ are nearly constant \citep{Zhang:2004ie}, and their derivatives can  be neglected.

\subsection{After the RS crossing time}

Let us denote $t_\times$ as the time when the RS crosses the ejecta shell. For $t > t_\times$, the FS and RS evolve independently.
The RS evolves according to the BM self-similar solution \citep{Gao:2015lga}.

For FS, the energy conservation reduces to:
\be
 {\rm d}E_2 = -\epsilon_2\gamma_2(\gamma_2-1)\,{\rm d}m_2c^2\,. 
\ee
Substituting Eq. (\ref{eq_E2}) to the left-hand side, we obtain:
\be
\frac{{\rm d}\gamma_2}{{\rm d}R} = -\frac{\gamma_2^2-1}{\left[\epsilon_2+2(1-\epsilon_2)\gamma_2\right]m_2} \frac{{\rm d}m_2}{{\rm d}R}\,. 
\label{eq_aftercross}
\ee
This is similar to the FS dynamical model in \citet{Huang:1999di}.  The only difference is the lack of $m_{\rm ej}$ terms in the denominator, since for $t > t_\times$ the ejecta mass will be insignificant ($m_2 \gg m_{\rm ej}$).

Notice after crossing, our following RS solutions are only valid for the relativistic stage as described by BM self-similar solutions.
The Newtonian stage should be described by ST self-similar solutions instead.
Since the late RS signals (e.g. $t\sim 10^5$ s or $\sim 1$ day) are much weaker than the FS, for most cases, there's no need to consider such a phase transition for the RS.
Either the BM or the ST solutions are not directly encoded in our model, and it's a free optional choice that could be applied in the numerical realization code.
Our FS solutions are analogous to \citet{Huang:1999di}, which is a modification of \citet{Piran:1999kx} focusing on enabling the transition to ST solutions.
In summary, after crossing our FS is valid for both the ultra-relativistic phase and the Newtonian phase, and our RS is currently focused on the relativistic phase.

\subsection{The hydrodynamical limit}

In the hydrodynamical limit ($\sigma \sim 0$),
the magnetization parameters become trivial:
$f_a \sim 1$,
$f_c \sim 1$.
Our model becomes greatly simplified:
\bea
Q &=& m_2 + m_3 + (1-\epsilon_2)(2\gamma_2-1)m_2\nonumber\\
  &&+(1-\epsilon_3)\left(2\gamma_{34}-\frac{u_4}{u_2}-1\right)m_3\,,
\\
P_2 &=& \gamma_2^2 - 1\,,
\\
P_3 &=& \gamma_2\gamma_{34} - \gamma_4\,.
\eea
Our new model Eq. (\ref{eq_result}) reduces exactly to the previous hydrodynamical model of \citet{1009-9271-7-6-05}:
\be
\frac{{\rm d}\gamma_2}{{\rm d}R} =
\frac{-4\pi R^2m_p\left[(\gamma_2^2-1)n_1 + \gamma_4(\gamma_2\gamma_{34}-1)(\beta_4/\beta_r-1)n_4\right]}
{\splitdfrac{m_2+m_3 + (1-\epsilon_2)(2\gamma_2-1)m_2}{+(1-\epsilon_3)(2\gamma_{34}-1-u_4/u_2)m_3}}\,.
\ee

\section{Dynamics Results}
\label{sec:result}

\subsection{Dynamics}
\label{sec:result_dyn}

\begin{figure*}
\centering
\includegraphics[width=0.99\textwidth]{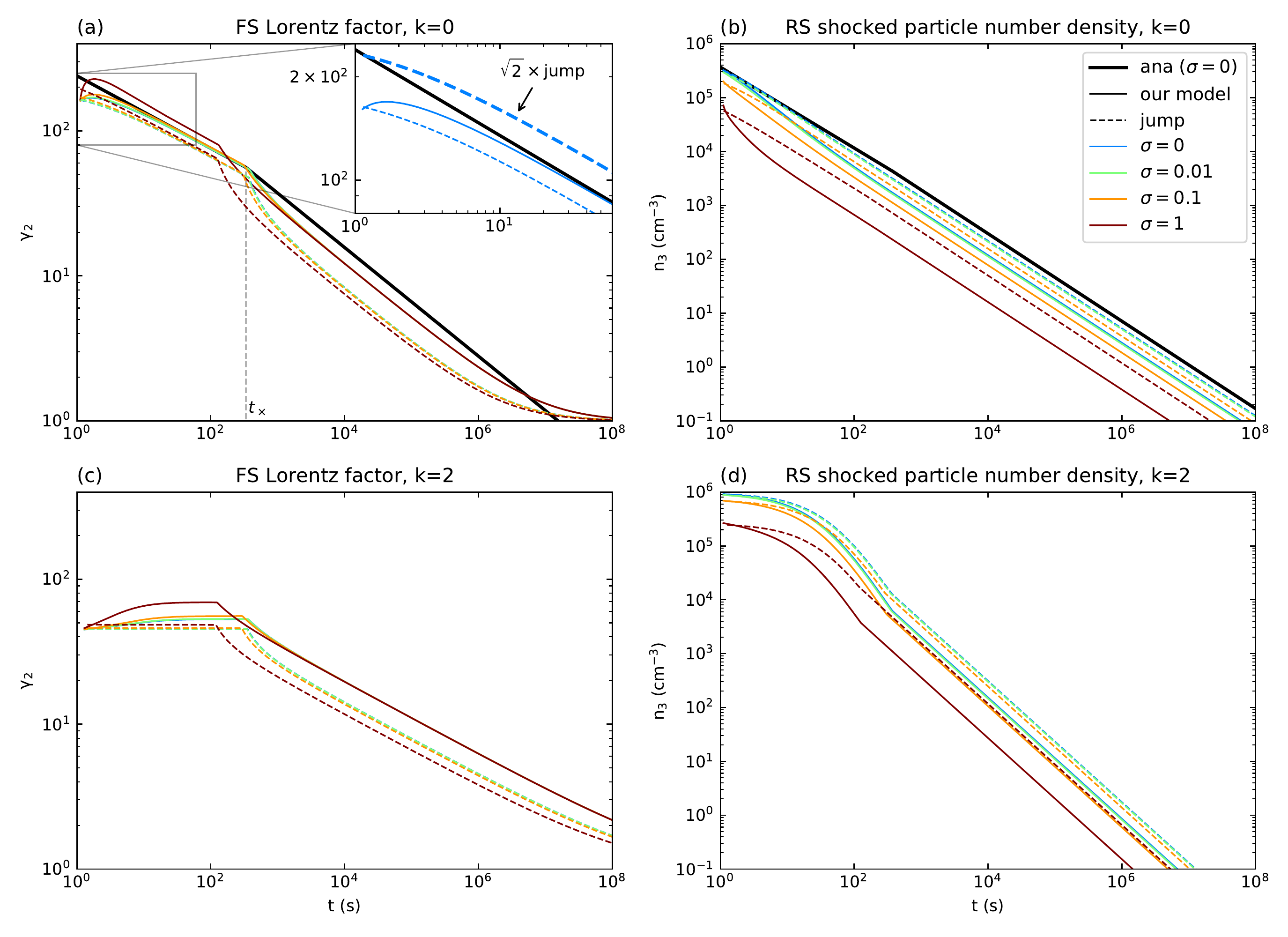}
\caption{
Dynamics results in the ISM cases ($k=0$, upper row), and the wind cases ($k=2$, bottom row).
The solid color lines show the results of our numerical model, and the dashed color lines show the results based on the shock jump condition model.
Black lines are analytical results in the hydrodynamical limit.
The left column shows the evolution of the bulk Lorentz factor $\gamma_2$ of the shocked CBM.
The turnings are the RS crossing times.
Panel (a) insert is a hydrodynamical comparison, where our model and the jump condition model differ with a factor $\lesssim\sqrt{2}$.
The right column shows the shocked ejecta particle number density $n_3$, where the jump condition model solutions are slightly larger. 
The deviation of $n_3$ becomes larger in higher $\sigma$ cases. 
}
\label{fig_dyn}
\end{figure*}

We numerically solved our dynamical model (Eq. \ref{eq_result}), in both ISM and wind cases, with the common setup parameters: $E_0 = 10^{52}\,{\rm erg}$, $\eta = 300$, $n_0 = 1~{\rm cm}^{-3}$ (ISM), $A_{\ast} = 0.01$ (wind), $\Delta_0 = 10^{13} \,{\rm cm}$, $z = 1$.
The $\Lambda$CDM cosmological parameters are $\Omega_m = 0.31$, $\Omega_\Lambda = 0.69$, and $H_0 = 68\,{\rm km~s^{-1}\,Mpc^{-1}}$.
We experimented with a range of magnetization values $\sigma = [0,0.01,0.1,1]$.

The upper row of Figure \ref{fig_dyn} shows the results of the dynamics for the ISM cases ($k=0$). 
In the bottom row, we present the results of a wind case ($k=2$).
To different magnetization, the Lorentz factor levels maintain small differences, except for the case $\sigma=1$, where $\gamma_2$ is larger than the rest cases, especially before the RS crossing.
Our Lorentz factor results are generally larger than that of the jump condition model.
For instance, comparing the model results in the hydrodynamical limit (shown in the insert to panel a), our Lorentz factor results are larger than the jump condition model results by a factor $\lesssim \sqrt{2}$.

With the increase of the $\sigma$ values, the particle number density of the shocked ejecta density $n_3$ is reduced.  In our model, the $n_3$ values are generally lower compared with the jump condition model solutions.  The reason is this: compared with the jump condition model, our model yields higher  $\gamma_{3}$,  higher  $\gamma_{34}$, and therefore lower $n_3$,   as determined by Eq. (\ref{eq_n3n4}).

\subsection{Energy conservation}
\label{sec_conser}

\begin{figure*}
\centering
\includegraphics[width=0.99\textwidth]{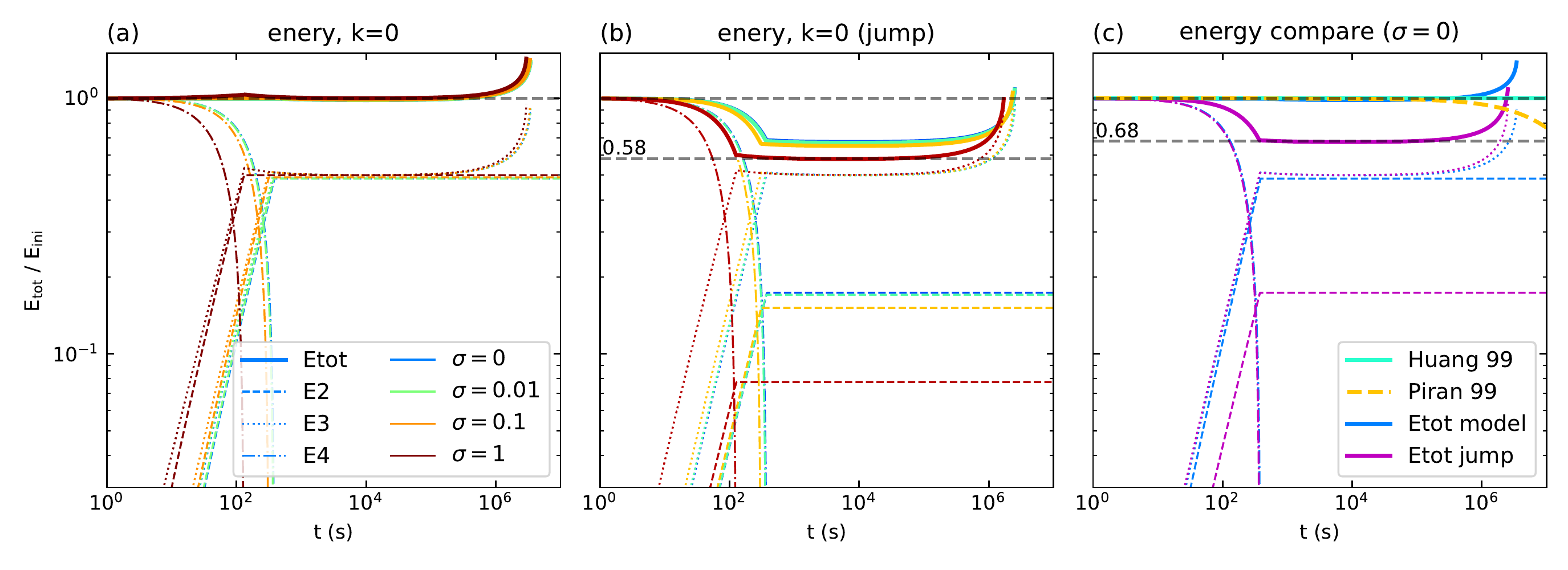}
\caption{
The energy conservation ($k=0$, and $\epsilon_2=\epsilon_3=0$).
Panel (a) is from our model.
Panel (b) is from shock jump condition model \citep{Zhang:2004ie}.
Panel (c) is the comparison of the hydrodynamical cases ($\sigma=0$), where we also include FS shock models from \citet{Piran:1999kx} (Piran 99) and \citet{Huang:1999di} (Huang 99).
Our model, Piran 99 and Huang 99 models are energy conserved.
The jump condition model continually loses energy in region 2.
Higher $\sigma$ case lose more energy, and by the time of RS crossing, the hydro case total energy drops to $68$\%, and the $\sigma=1$ case drops to $58$\%.
At the end of these evolutions, some models suffer from either increase or drop of the total energy due to not satisfying the ST self-similar solutions.
Since we focused on the relativistic stage, the RS results (both our model and jump condition model that we reproduced) follow the BM self-similar solutions and lack a transition to the ST self-similar solutions in the Newtonian stage.
This makes an increase of total energy when $t>10^5$ s. 
As for the Piran 99 result of the FS, it doesn't satisfy the ST self-similar solutions and leads to a loss in the total energy in the Newtonian stage.
}
\label{fig_conserv}
\end{figure*}

In Figure \ref{fig_conserv} we compare the conservation accuracy of our model and the jump condition model in the ISM case ($k=0$).
These numerical results are all in the adiabatic regime so that no energy is supposed to lose throughout the whole evolution either in the shocked CBM (i.e. $\epsilon_2=0$) or in the shocked ejecta (i.e. $\epsilon_3=0$). 

Our model results presented in panel (a)  demonstrate the conservation of the total energy. Initially, all energy is contained in the ejecta (region 4).  During the dynamical interaction, this energy is flowing into the shocked ejecta (region 3) and the shocked CBM (region 2). By the time of the $t_\times$, this energy transfer process is complete. The interesting phenomenon is that the energies contained in regions 2 and 3 are almost equal by the time of RS crossing, which is roughly 50\% of the total energy each.

Panel (b) shows that the jump condition model loses a significant fraction of energy from region 2, while region 3 doesn't.  In the jump condition model, the Lorentz factor decreases by a factor $\lesssim\sqrt{2}$.  The decrease in the value of $\gamma_2=\gamma_3$ makes $n_2\propto\gamma_{21}$ lower, but $n_3\propto\gamma_{34}$ higher.  This is the reason why only the FS shocked region loses energy in the jump condition model, while in the RS shocked region the decrease of the Lorentz factor is recompensed by the increase of particle number density.
In the hydrodynamical ISM case, energy losses of the jump condition model reach $32$\%.  With the increase in the value of $\sigma$, more energy losses. In the $\sigma=1$ case, the total energy loses $42$\% by the time of RS crossing. 

The wind cases are similar to the ISM cases.
In the wind cases, the energy losses are $25$\% in the hydrodynamical limit and $38$\% for $\sigma=1$, slightly lower than in the ISM case.

Panel (c) compares the hydrodynamical cases, including two single-shock models \citet{Piran:1999kx}  and \citet{Huang:1999di} , which are also energy conserving.

Worth mentioning that our model and our reproduction of the jump condition model ignored the RS transition for ST solutions in the Newtonian stage, hence at late time $t>10^5$ s the RS energy is increased a bit as reflected in the figure.
The increase is caused by the region 3 internal energy $\propto\gamma_3(\gamma_{34}-1)$ (in source frame, see Eq. \ref{eq_E3}).
When $\gamma_3\rightarrow 1$, the product of $\gamma_3(\gamma_{34}-1)$ blows up.
However, this will not harm the lightcurves much, since at that time FS is dominated and RS lightcurves will be buried.
On the other hand, the \citet{Piran:1999kx} model also fails in transiting to the ST solutions.
As pointed out by \citet{Huang:1999di}, the ST solutions require $\beta\propto R^{-3/2}$, but \citet{Piran:1999kx} model yields $\beta\propto R^{-3}$.
This means \citet{Piran:1999kx} model has a loss in the total energy of the Newtonian stage.
This problem will be more significant since this happens in the FS.

\subsection{Pressures}
\label{sec:pressure}

\begin{figure*}
\centering
\includegraphics[width=0.99\textwidth]{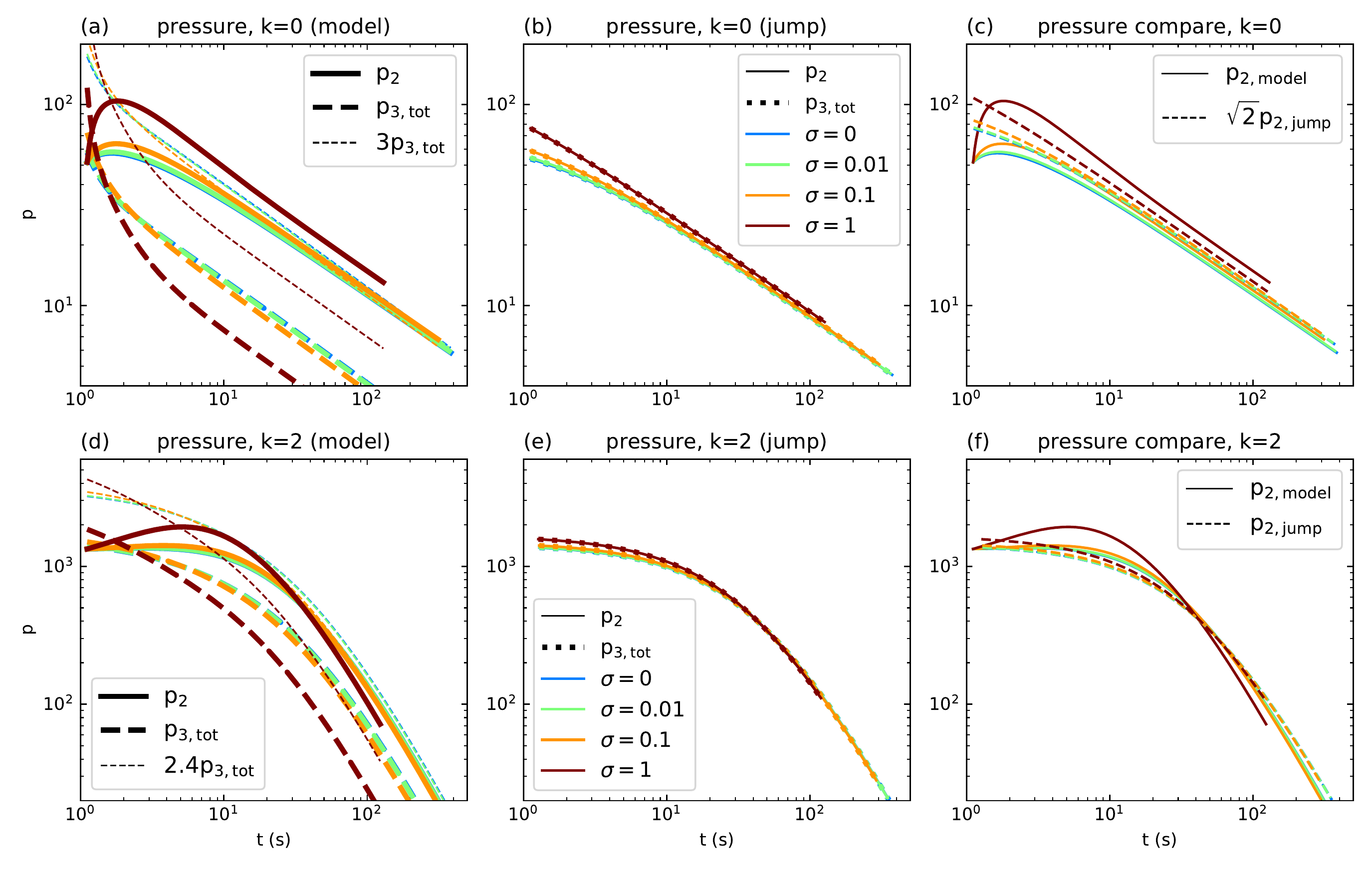}
\caption{
The pressures comparison before the RS crossing.
Panel (a) is from our model, by the end of the crossing satisfying $p_2\simeq 3p_{\rm 3,tot}$ (except $\sigma=1$ case).
Panel (b) is from the jump condition model, where uniform pressure is strictly imposed $p_2= p_{\rm 3,tot}$, which is questioned for conflicting the total energy conservation.
Panel (c) is the comparison between our model and the jump condition model, a deviation $p_2/p_{\rm 2,jump}\approx\sqrt{2}$ is seen in the FS shocked region (except $\sigma=1$ case).
Bottom row are wind cases ($k=2$), where $p_2/p_{\rm 3,tot}\approx 2.4$ and  $p_2/p_{\rm 2,jump}\approx 1$.
}
\label{fig_pressure}
\end{figure*}

The mechanic model proposed by \citet{2006ApJ...651L...1B} is energy-conserving since it includes constraints on the total energy conservation.
In their example wind case ($k=2$), the pressures satisfy $p_f = 3p_r$, where $p_f\simeq p_2$ and $p_r\simeq p_3$ are pressures at the  FS and RS fronts, respectively.
We show that our model achieves similar pressure relations.

In Figure \ref{fig_pressure} we compare pressure relations of our model and also the jump condition model before the RS crossing. 
Panel (a) shows that our initial values satisfy the pressure equality, but they are quickly driven to an asymptotic relation that satisfies $p_2\simeq 3p_{\rm 3,tot}$, where $p_{\rm 3,tot} = p_3+p_{\rm B,3}$. The wind cases in panel (d) roughly satisfy $p_2\simeq 2.4p_{\rm 3,tot}$. 
The exceptions are when $\sigma=1$ for both the ISM and wind cases, where the deviations of the pressures in the two regions become larger, mainly due to the behavior of $n_3$ shown in Figure \ref{fig_dyn}.  

Panels (b) and (e) show that the shock jump condition model is strictly governed by the pressure uniformity assumption. 
Panel (c) shows the comparison of our energy-conserving model with the jump condition model for the ISM cases, the FS shocked region pressure $p_2$ is larger by a factor around $\sqrt{2}$ than the jump condition model.
For the wind cases, the two models show similar levels of FS pressures, as shown in Panel (f).

\section{Synchrotron emissions}
\label{sec:syn}

\subsection{Magnetic energy fraction and spectral parameters}
\label{sec:frac}

\begin{figure*}
\centering
\includegraphics[width=0.99\textwidth]{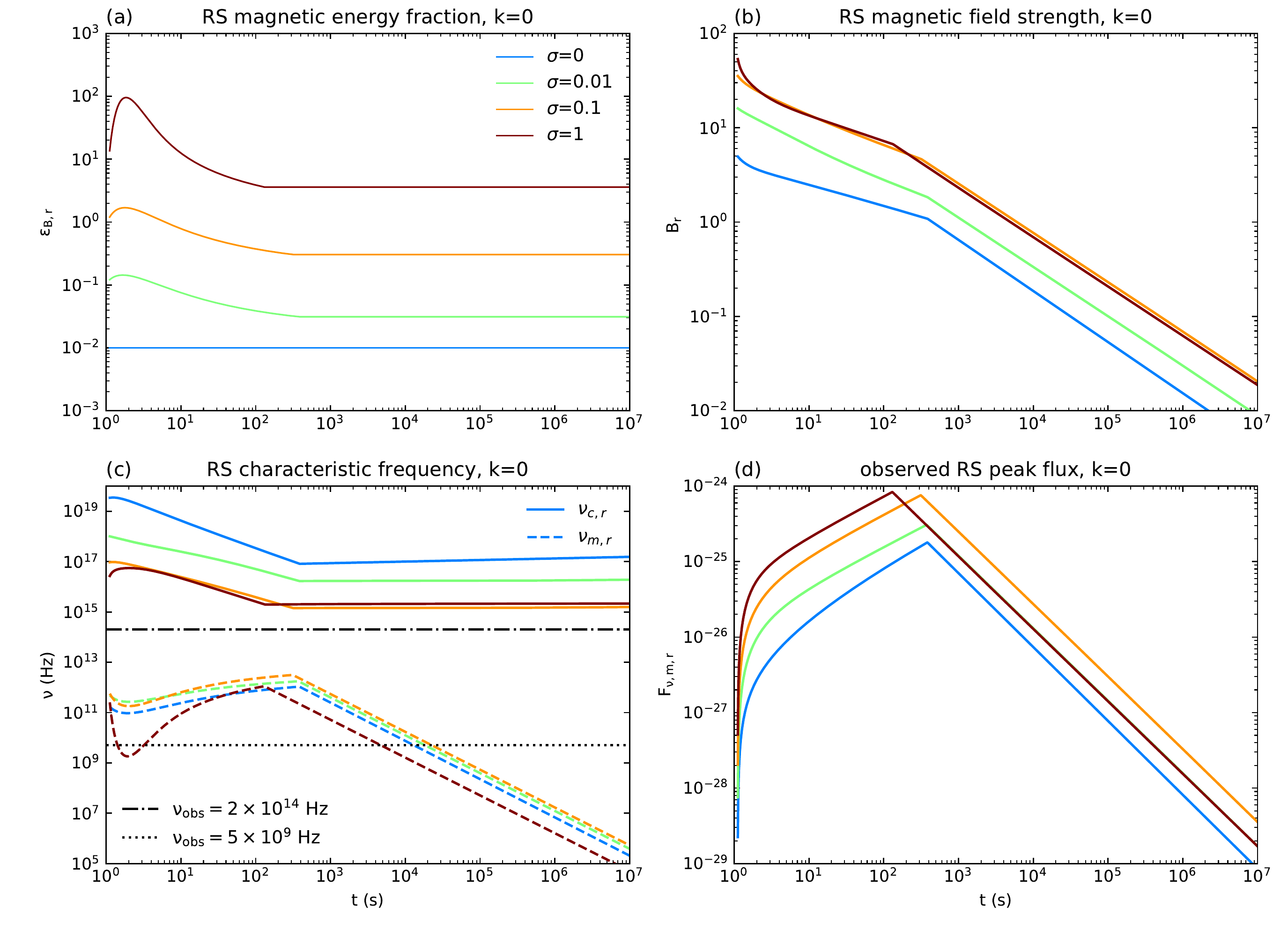}
\caption{
The spectral related parameters of the RS shocked region ($k=0$).
Panel (a) is the magnetic energy density to thermal energy density ratio $\epsilon_{B,r}$ defined in Eq. (\ref{eq_eps}).
Panel (b) is the magnetic field strength $B_r$.
Panel (c) presents characteristic frequencies $\nu_{c}$ and $\nu_{m}$, and the observational optical and radio frequencies that we used in deriving the light curves. 
These are all of slow cooling cases since $\nu_c> \nu_m$.
Panel (d) is the peak spectral power $F_{\nu,m}$.
The magnitudes of $B_r$ and $F_{\nu,m}$ are dependent to the magnetization $\sigma$, their maximum levels are at $\sigma\sim 0.1-1$.
}
\label{fig_eps}
\end{figure*}

Based on the results of the dynamics, synchrotron emission light curves can be obtained \citep{Sari:1997qe}.
The electrons are accelerated into a power-law distribution with an index that we adopt as $p = 2.5$.
We assume a constant fraction $\epsilon_{e}$ of the shock energy goes into electrons, for FS and RS shocked regions $\epsilon_{e,f} = \epsilon_{e,r} = 0.1$.

To discuss the synchrotron emission from the hydrodynamical CBM where $\sigma=0$ seems to be counter-intuitive.
Though the initial magnetic field in the shocked regions is usually negligible, but certain plasma instabilities, e.g. the current-driven instability \citep{Reville_2006}, the Kelvin-Helmholtz shear instability \citep{Zhang_2009}, the Weibel/filamentation instability \citep{Medvedev_1999,10.1111/j.1365-2966.2009.15869.x,Tomita_2016}, the \v{C}erenkov resonant instability \citep{10.1111/j.1365-2966.2009.15869.x}, the Rayleigh-Taylor instability \citep{Duffell_2013}, the magneto-rotational instability \citep{Cerd_Dur_n_2011}, etc., or the pile-up effect \citep{10.1093/mnras/stu2104} could generate and amplify the magnetic field which is essential to the particle accelerations.
For simplicity reason, we ignore the details of the magnetic field amplification process, and assume in the CBM the magnetic energy density is a constant fraction $\epsilon_{B,f} = 0.01$ of the thermal energy density \citep{Sari:1997qe,Zhang:2004ie}.

As for the RS shocked region, for $\sigma=0$ case we assume similar field amplification processes with the FS shocked region, so $\epsilon_{B,r}\sim\epsilon_{B,f}$.
For the $\sigma>0$ cases of the RS shocked region, on the contrary, for simplicity, the field amplification is ignored.
Before the RS crosses the ejecta, the RS magnetic energy fraction is inferred from Eq. (\ref{eq_eb3e3}).
After the RS crosses the ejecta, the definition of factor $f_c$ is invalid, we use dummy values by suspending the crossing time value.
Despite the potential limitations of the simplification, our RS magnetic energy fraction is finally assumed as (see panel (a) in Figure \ref{fig_eps}):
\be
\epsilon_{B,r}
=
\begin{cases}
\epsilon_{B,f} & \text{if $\sigma=0$}\,,
\\
(\hat\gamma_3-1)(f_c-1) & \text{if $\sigma> 0$ and $t\le t_\times$}\,,
\\
(\hat\gamma_3-1)(f_c(t_\times)-1) & \text{if $\sigma> 0$ and $t>t_\times$}\,.
\end{cases}
\label{eq_eps}
\ee

Once we know the magnetic energy density, the corresponded magnetic field strength $B$ is also known from the relation (see panel (b) in Figure \ref{fig_eps}) \citep{Sari:1997qe}:
\be
\frac{B_r^2}{8\pi} = e_3\epsilon_{B,r}\,.
\label{eq_Br}
\ee

In Figure \ref{fig_eps}, panel (a) shows $\epsilon_{B,r}$ increases when $\sigma$ increases.
Panel (b) shows $B_r$ reaches a peak level at around $\sigma\sim 0.1-1$, this is because $e_3$ (or equivalently $n_3$, right column panels in Figure \ref{fig_dyn}) decreases with the increasing of $\sigma$.
Panel (c) shows the relation between minimum frequency $\nu_m$, critical frequency $\nu_c$ and chosen observational frequencies $\nu_{\rm obs}$ for the RS.
The frequency relation determines the light curve power-law scalings according to the cooling rules to be either fast or slow coolings \citep{Sari:1997qe}.
Our cases in Panel (c) are all of slow coolings since $\nu_c>\nu_m$.
Panel (d) shows the observed peak flux $F_{\rm\nu,m,r}$ reach to a maximum level at $\sigma\sim 0.1-1$ just like the levels of $B_{r}$.

\subsection{Optical light curves}
\label{sec:optic}

\begin{figure*}
\centering
\includegraphics[width=0.99\textwidth]{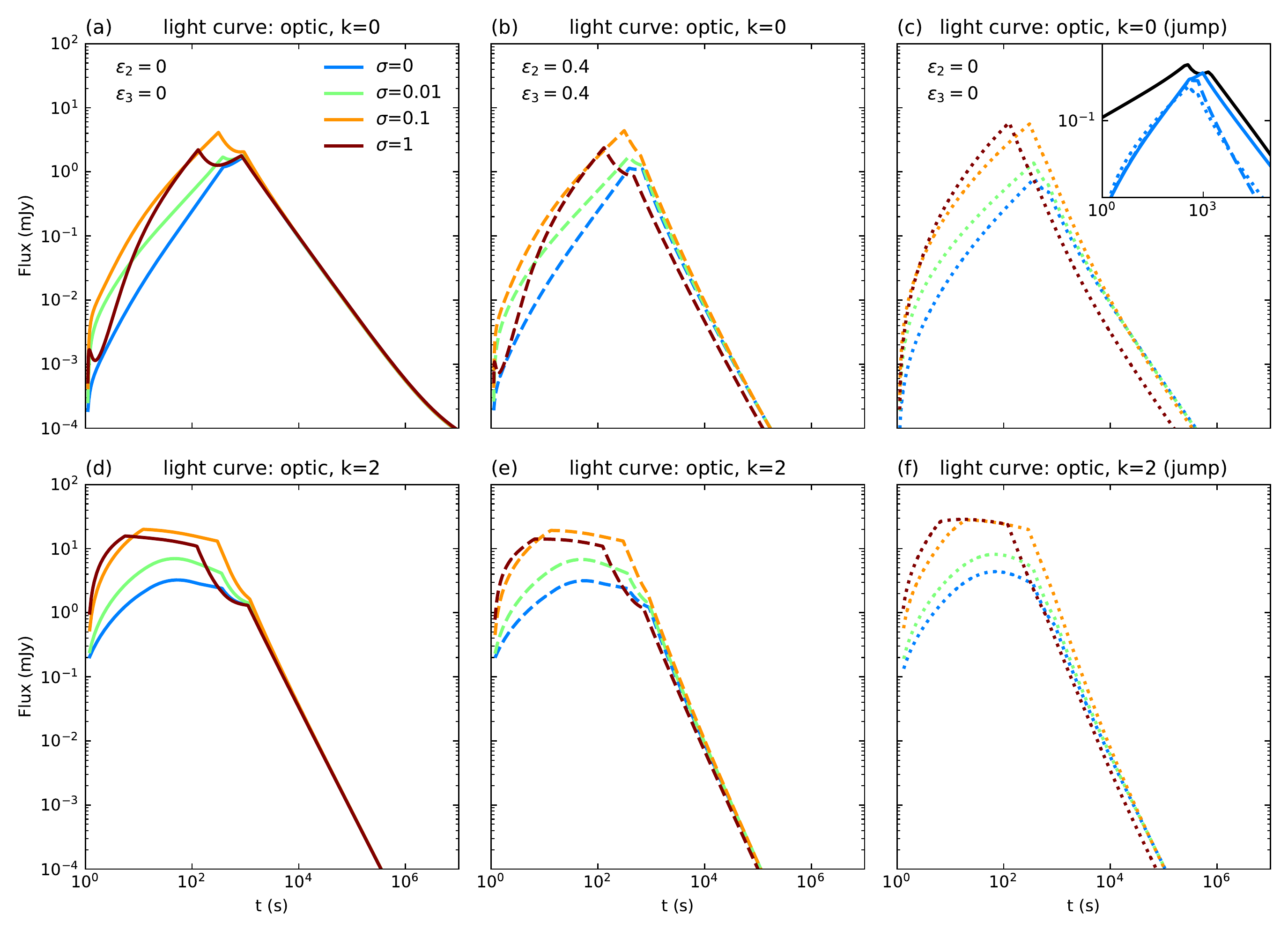}
\caption{
Total light curves for optic frequency $2\times 10^{14}$ Hz.
In the upper row are the ISM cases ($k=0$), and at the bottom are the wind cases ($k=2$). 
The left column panels are the light curves of the adiabatic cases of our model.
The middle column panels are the light curves of the radiative cases with $\epsilon_2=\epsilon_3=0.4$ of our model.
The right column panels are light curves derived from the magnetized shock jump condition model \citep{Zhang:2004ie}.
Panel (c) insert is a hydro case comparison.
Our model adiabatic results differ with the jump condition model solutions especially in the late emission (mainly from FS), since the jump solutions unexpectedly lose energy and become similar to our radiative cases.
}
\label{fig_flux}
\end{figure*}

In Figure \ref{fig_flux}, we present the optical light curves, with observed frequency $\nu_{\rm obs}=2\times 10^{14}\,{\rm Hz}$.
The jump condition model is not implemented with radiative terms, so is only applied to the adiabatic cases.
Compared with the adiabatic cases in our model, the late time signals in the jump solutions are smaller.
The jump solutions are comparable to our model results with radiative efficiencies $\epsilon_2=\epsilon_3 = 0.4$, which lose $40$\% of the thermal energy.

Panel (c) insert is a hydro case comparison.
For the early time light curves, the jump condition model result is similar to our model result.
The analytical light curve is exaggerated caused by the distortion of the approximations at the early time.
For the late-time light curves, the jump condition model result is suppressed due to energy loss in region 2 and agrees with a radiative case in our model.

All the cases shown in Figure \ref{fig_flux} reach to peak levels at around $\sigma\sim 0.1-1$ \citep{Zhang:2004ie,2004A&A...424..477F}.
The early signals (mainly from RS) are sensitive to the ejecta magnetization, which indicates that the early observational light curves are possible sources to limit the ejecta magnetization $\sigma$.
The dependence of the light curves on ejecta magnetization in the optical bands lasts in the timescale of hours ($\sim 10^3$ s).

\subsection{Radio light curves}
\label{sec:radio}

\begin{figure*}
\centering
\includegraphics[width=0.99\textwidth]{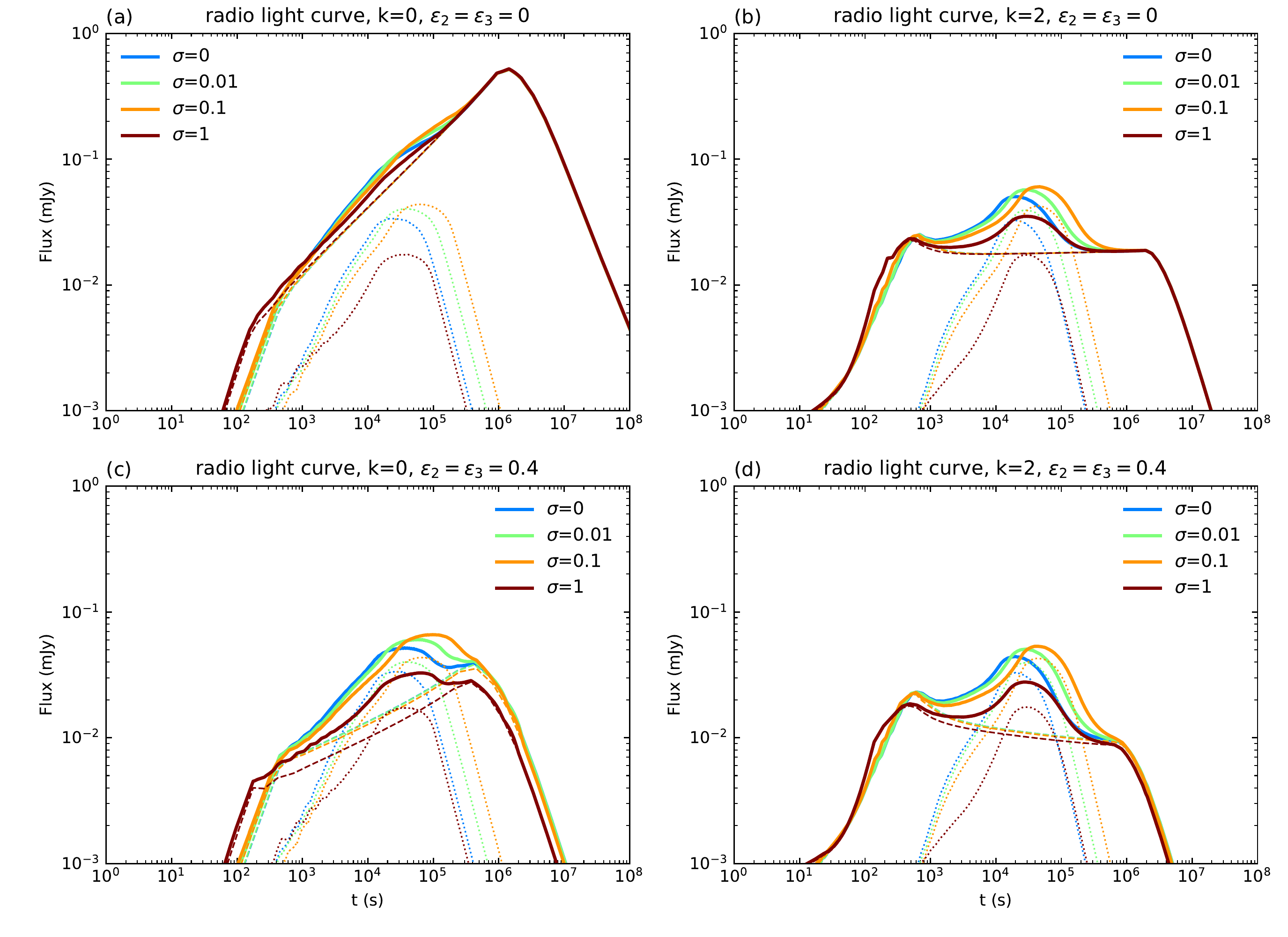}
\caption{
Total light curves for radio band with observational frequency $5\times 10^9$ Hz.
The synchrotron self-absorption effect is included. 
The dynamics setup is the same as Figure \ref{fig_flux}.
The magnitude of $\sigma$ affects dramatically the RS light curves (dotted lines).
The FS light curves are affected by $\sigma$ slightly in the early time, but only the radiative cases can also affect the late time emissions.
The peaked times are around 1 day ($\sim 10^5$ s), and the dependence of the light curves on $\sigma$ lasts to a timescale of a month ($\sim 10^6$ s).
}
\label{fig_flux_radio}
\end{figure*}

The radio light curves are shown in Figure \ref{fig_flux_radio} where $\nu_{\rm obs} = 5\,{\rm GHz}$.
We include the synchrotron self-absorption (SSA) effect, which is important to the low frequencies.
The RS light curves are sensible to the magnitude of ejecta magnetization and the signatures of $\sigma$ can last around a month ($\sim 10^6$ s).
The late-time light curves are dominated by the FS and can last for even a year ($> 10^7$ s), but they are insensitive to $\sigma$, except some minor variations may be observed in the radiative cases.

\subsection{Model test: GRB181201A}
\label{sec:GRB181201A}

\begin{figure}
\centering
\includegraphics[width=0.49\textwidth]{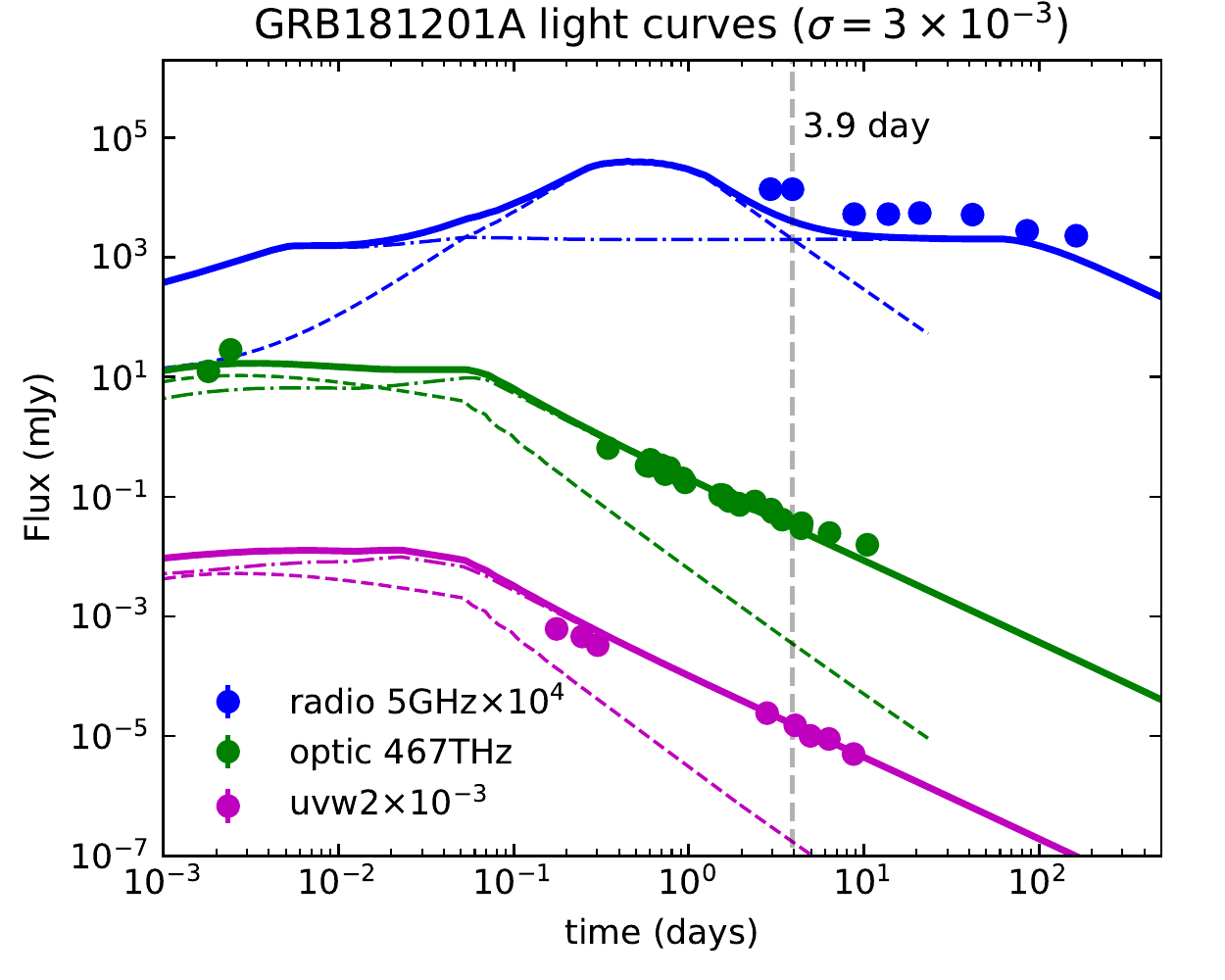}
\caption{
Our model results compared with the observational light curves data of GRB181201A.
The dashed lines are from the RS, dash-dotted lines are from the FS, and the solid lines are the total light curves.
The most important parameters are suggested by \citet{Laskar_2019_1} as:
$E_{\rm K,iso} \approx 2.2\times 10^{53}$ erg,
$p\approx 2.1$,
$\epsilon_e \approx 0.37$,
$\epsilon_B \approx 9.6\times 10^{-3}$,
$A_\ast \approx 1.9\times 10^{-2}$,
$\sigma= 3\times 10^{-3}$.
We choose a shell width as $\Delta_0 = 1\times 10^{14}$ cm.
In this burst, the unique features from the RS are the breaking in the radio light curve at $3.9$ days and the very early optical flare.
}
\label{fig_GRB181201A}
\end{figure}

We test our model by comparing our theoretical light curves to the multi-wavelength observations of GRB181201A.
\citet{Laskar_2019_1} fitted GRB181201A by the Markov Chain Monte Carlo (MCMC) method using the python package {\tt emcee} \citep{Foreman_Mackey_2013}, derived the most likely parameters from the multidimensional parameter space as:
$E_{\rm K,iso} \approx 2.2\times 10^{53}$ erg,
$\eta= 103$,
$p\approx 2.1$,
$A_\ast \approx 1.9\times 10^{-2}$,
$\epsilon_e \approx 0.37$,
$\epsilon_B \approx 9.6\times 10^{-3}$,
$\sigma \approx 3\times 10^{-3}$,
$z=0.45$,
$\Omega_m = 0.31$, 
$\Omega_\Lambda = 0.69$,
$H_0 = 68\,{\rm km~s^{-1}\,Mpc^{-1}}$.
Besides, we adopt a shell width with $\Delta_0 = 1\times 10^{14}$ cm.
Based on these parameters, we compare the light curves simulated from our model with the observational data, in radio, optical, and UV bands in Figure \ref{fig_GRB181201A}.
Our model successfully re-created the radio light curve turning at 3.9 days and the early optical rise, which are key signatures from the RS \citep{Laskar_2019_1}.

\section{Discussions and Conclusions}
\label{sec:conc}

The GRB afterglows can be described by the shock jump conditions.
To the FS-RS system, one common approach of solving the jump conditions is based on additional equations that assuming the whole shocked region has uniform pressure and Lorentz factor \citep{1995ApJ...455L.143S,2000ApJ...545..807K,2004A&A...424..477F,Fan:2004fe,Zhang:2004ie}.
However, the uniform pressure assumption causes an energy conservation problem \citep{2006ApJ...651L...1B}.
The uniform Lorentz factor assumption also harms the energy conservation, but for the sake of reducing the complication, it is usually preserved, while the energy conservation is reassured by a total energy conservation equation.
To study the GRB afterglows from the magnetized ejecta, we extend a hydrodynamical energy-conserving FS-RS dynamical model \citep{1009-9271-7-6-05} to the MHD limit by embedding magnetized shock jump conditions \citep{Zhang:2004ie}.
In detail, before RS crosses the ejecta shell, we replace the uniform pressure assumption with a total energy conservation equation.
After the RS crosses the ejecta shell, the RS evolves self-similarly, while the FS continues to be decelerated by the CBM governed by a hydrodynamical FS model \citep{Huang:1999di}.

The results of Lorentz factors in our new model are larger by a factor $\lesssim\sqrt{2}$ compared with the jump condition model solutions.
Before the RS crosses the ejecta shell, the total pressures of the shocked ejecta and the shocked CBM are no longer equal, and gradually reach an asymptotic relation similar to the description of the mechanic model \citep{2006ApJ...651L...1B}.
When $\sigma<1$, our ISM cases follow $p_2\simeq 3p_{\rm 3,tot}$, and wind cases $p_2\simeq 2.4p_{\rm 3,tot}$.

The shock jump condition model \citep{Zhang:2004ie} is an adiabatic model, in which we find the ISM cases lose around $32-42$\% of the total energy when $\sigma\le 1$, and the wind cases $25-38$\%.
Cases with larger $\sigma$ lose more energy.
The energy loss happens only in the FS shocked region, making the late-time light curves change corresponded to the ejecta magnetization, and is comparable to the radiative mode in our model.

The RS in the shock jump condition model doesn't suffer from the energy conservation problem, since the reduction of the Lorentz factor is recompensed by the amplification of the particle number density.
Our model and the jump condition model have similar behaviors in the RS radiation -- at first, the observed luminosity increases together with increasing $\sigma$, then the peak luminosity is achieved when $\sigma \sim 0.1-1$.
Once $\sigma > 1$, the luminosity declines because the total number density of the shocked ejecta begins to drop significantly \citep{Zhang:2004ie,2004A&A...424..477F}. 
The imprint of $\sigma$ on the RS is obviously in the early time light curves, for optical bands it lasts with a timescale of hours ($10^3$ s), but for radios, it lasts up to a month ($10^5$ s).
The dependence of early emissions on $\sigma$, making it possible to constrain the magnetization of the ejecta by early observational light curves.

\section*{Acknowledgments}

Special thanks to Krzysztof Nalewajko, who helps us a lot during the process of this project, especially in deriving and numerically solving the shock jump conditions, verification of the calculations, and also for editing the texts.
Thank the anonymous reviewer for the constructive suggestions, like the RS existence conditions, radio emissions, the comparison to the observations, etc., and for much other detailed feedback.
This work is supported by the Polish National Science Center grant 2015/18/E/ST9/00580.

\section*{DATA AVAILABILITY}

The data underlying this article will be shared on reasonable request to the corresponding author.

\bibliographystyle{mnras}

\bsp	
\label{lastpage}
\end{document}